\documentclass{PoS}

\usepackage{mathptm}  

\newcommand{\beq}{\begin{eqnarray}}
\newcommand{\eeq}{\end{eqnarray}}
\newcommand{\be}{\begin{eqnarray*}}
\newcommand{\ee}{\end{eqnarray*}}

\newcommand{\eg}{{\it e.g.}}
\newcommand{\etal}{{\it et al.}}

\renewcommand{\L}{\mathcal L}

\newcommand{\glabcms}{\gamma^{\rm lab}_{\rm cms}}
\newcommand{\blabcms}{\beta^{\rm lab}_{\rm cms}}

\def\lsim{\raise0.3ex\hbox{$<$\kern-0.75em\raise-1.1ex\hbox{$\sim$}}}
\def\gsim{\raise0.3ex\hbox{$>$\kern-0.75em\raise-1.1ex\hbox{$\sim$}}}

\def\jpsi    {\mbox{$J/\psi$}}

\def\beq     {\begin{equation}}
\def\eeq     {\end{equation}}

\title{A Fixed-Target ExpeRiment at the LHC (AFTER@LHC) : luminosities, 
target polarisation and a selection of physics studies}

\ShortTitle{AFTER @ LHC}

\author{\speaker{J.P.~Lansberg}, V.~Chambert, J.P.~Didelez, B.~Genolini, C.~Hadjidakis, P.~Rosier
        \\
        IPNO, Universit\'e Paris-Sud, CNRS/IN2P3, F-91406, Orsay, France
}

\author{R.~Arnaldi, E.~Scomparin\\
        INFN Sez. Torino, Via P. Giuria 1, I-10125, Torino, Italy
        }
\author{S.J.~Brodsky\\
        SLAC National Accelerator Laboratory, Stanford University, Menlo~Park, CA 94025, USA
        }

\author{E.G.~Ferreiro\\
        DFP \& IGFAE, Universidade de Santiago de Compostela, 15782 Santiago de Compostela, Spain
        }

\author{F.~Fleuret\\
        Laboratoire Leprince Ringuet, \'Ecole Polytechnique, CNRS/IN2P3,  91128 Palaiseau, France
        }

\author{A.~Rakotozafindrabe\\
        IRFU/SPhN, CEA Saclay, 91191 Gif-sur-Yvette Cedex, France
        }

\author{U.I.~Uggerh\o j\\
        Department of Physics and Astronomy, University of Aarhus, Denmark
        }

\abstract{We report on a future multi-purpose fixed-target experiment
with the proton or lead ion LHC beams extracted by a bent crystal. The multi-TeV LHC
 beams  allow for the most energetic fixed-target experiments ever performed.
Such an experiment, tentatively named AFTER for "A Fixed-Target ExperRiment", gives access to new domains of particle and 
nuclear physics complementing that of collider experiments, in particular at RHIC and at the EIC projects. The 
instantaneous luminosity at AFTER using typical targets surpasses that of RHIC by more than 3 orders 
of magnitude. Beam extraction by a bent crystal offers an ideal 
way to obtain a clean and very collimated high-energy beam, without 
decreasing the performance of the LHC. The fixed-target mode also has the advantage
of allowing for spin measurements with a polarised target and for an access over the full 
backward rapidity domain up to $x_F \simeq - 1$. Here, we elaborate on the reachable 
luminosities, the target polarisation and a selection of measurements with hydrogen and deuterium targets.
}

\FullConference{Sixth International Conference on Quarks and Nuclear Physics\\
		 April 16-20, 2012\\
		 Ecole Polytechnique, Palaiseau,  Paris}

\begin{document}

\section{Introduction}

The important contributions of fixed-target experiments to hadron and nuclear physics, 
especially in accessing the high Feynman $x_F$ domain and in offering a number of options 
for polarised and unpolarised proton and nuclear targets need not be recalled.
For those who are not convinced, let us simply recall that they have 
led to particle discoveries ($\Omega^-(sss)$, $J/\psi$, $\Upsilon$,...) as well as
evidence for the novel dynamics of quarks and gluons in heavy-ion collisions. 
They  have also led to the observation of surprising QCD phenomena: the 
breakdown of the Lam-Tung relation,  colour transparency, higher-twist effects at high 
$x_F$, anomalously large single- and double-spin correlations, 
and the breakdown of factorisation in $J/\psi$ hadroproduction at high 
$x_F$ in proton-nucleus collisions (see~\cite{Brodsky:2012vg} and references therein).

High luminosities can be reached at fixed-target experiments thanks to the density and length of the target.
A wide spectrum of precision measurements at laboratory energies never reached before 
can be carried out thanks to the LHC beams of 7 TeV protons and 2.76 TeV-per-nucleon lead ions interacting on a 
fixed-target. In addition, an entire set of heavy hadrons such as the $\Omega^{++} (ccc)$ and exotic states
could be looked at with a unique access to the large negative-$x_F$ domain. 

7 TeV protons colliding on fixed targets release  a center-of-mass 
energy close to 115 GeV, in a range never explored thus far, between those of SPS and RHIC. 
With the benefit of the proton runs lasting nine months each year, the production of quarkonia, open heavy flavour hadrons
and prompt photons in  $pA$ collisions can thus be investigated with statistics previously unheard of and in 
the backward region, $x_F < 0$, which is essentially uncharted.
In addition to conventional targets of Pb, Au, W, Cu, etc., high precision QCD measurements 
can also obviously be carried out in $pp$ and $pd$ collisions with hydrogen and deuterium targets.

For instance, at large negative $x_F$, intrinsic heavy quark distributions can be investigated by 
looking at new mechanisms for the production of hadrons with multiple heavy quarks such 
as  baryons with two or three bottom quarks as well as systems such as $\jpsi+D$~\cite{Lansberg:2012kf}. A deuterium target and gluon sensitive probes 
--such as quarkonium, open heavy-flavour and prompt-photon production-- will allow for a 
first precision measurement of the gluon content in the neutron.
 Polarizing the target would allow one 
to study spin correlations including the Sivers effect 
beyond conventional factorisation; this effect pins down the correlation between the parton $k_T$ and the nucleon spin. In particular, AFTER can bring much information on the contribution of the gluon angular momentum to the nucleon spin.

\section{Luminosities for a bent-crystal-extraction mode on the LHC proton beam}
\label{sec:lumi}

The idea of extracting a small fraction of the CERN LHC beam to be used for 
fixed target physics is not new. Already, in the early 90's, the LHB 
collaboration submitted a letter of intent to the appropriate committee~\cite{LHCB} 
to get an experiment based on bent-crystal extraction approved. At the time, this 
idea was turned down, mainly with the justification that the irradiation limit for 
the degradation of channeling performance was only known to be higher than $10^{19}$ 
particles/cm$^2$, and \emph{expected} to be up to $10^{22}$ particles/cm$^2$. 
Experiments have meanwhile shown that the degradation is approximately $6\%$ per 
$10^{20}$ particles/cm$^2$, see \eg~\cite{Baur00}. For realistic impact parameters 
and beam sizes at the crystal location, this corresponds to about one year of 
operation, after which the crystal has to be moved less than a millimeter to let 
the beam impact on an intact spot, a procedure that can be repeated almost at will.

Following this, and other important developments in the field of channeling in bent 
crystals, the Large Hadron Collider Committee now expresses that "...it may well be 
feasible to bend even the low emittance LHC beam" and that "possible future applications 
of the bent-crystal scheme abound: including beam-halo cleaning and \emph{slow extraction}" 
[our emphasis]. Thus, the committee recommends further studies to be performed at the 
LHC~\cite{LHCC107}.

One possibility that deserves to be studied further is the proposal~\cite{Uggerhoj:2005xz} 
to "replace" the kicker-modules in LHC section IR6 (the beam dump) by a bent crystal that 
will provide the particles in the beam halo with the sufficient kick to overcome the septum 
blade and to be extracted. By this method, a beam of about $5\cdot10^8$ protons/s can be 
extracted in the direction of the dump, at an expense of practically zero - the beam halo 
has to be removed by collimation anyway. Another possibility would be to integrate the 
extraction in a "smart collimator" solution, originally proposed by Valery Biryukov~\cite{Biry03}. 
This is presently the route followed by the CERN LUA9 collaboration. 

We have summarised the instantaneous luminosities which can be reached 
with various 1cm thick targets in table~\ref{tab:lumi-pA}. The
integrated luminosities over one year (taken as 10$^7$~s for the proton beam) are 
also given;  it is of the order of a fraction of a femtobarn$^{-1}$. Luminosities for the 
Pb run can be found in~\cite{Brodsky:2012vg}. Note that 1m long targets of liquid hydrogen or deuterium
give luminosities close to 20 femtobarn$^{-1}$.

\begin{table}[!hbt]
\centering\setlength{\arrayrulewidth}{.8pt} 
\renewcommand{\arraystretch}{0.9}\scriptsize
\begin{tabular}{|ccccc|ccccc|}
\hline
Target       & $\rho$        &$A$ & $\L$                     & $\int dt\L$ 
&Target       & $\rho$        &$A$ & $\L$                     & $\int dt\L$\\
(1 cm thick) & (g cm$^{-3}$) &    & ($\mu$b$^{-1}$ s$^{-1}$) & (pb$^{-1}$ yr$^{-1}$)&
(1 cm thick) & (g cm$^{-3}$) &    & ($\mu$b$^{-1}$ s$^{-1}$) & (pb$^{-1}$ yr$^{-1}$)\\
\hline\hline 
solid H  & 0.088 & 1   & 26 & 260 &
liquid H & 0.068 & 1   & 20 & 200 \\
liquid D & 0.16  & 2   & 24 & 240 &
Be       & 1.85  & 9   & 62 & 620 \\
Cu       & 8.96  & 64  & 42 & 420 &
W        & 19.1  & 185 & 31 & 310 \\
Pb       & 11.35 & 207 & 16 & 160 &
  &  &  &  &  \\
\hline
\end{tabular}
\caption{Instantaneous and yearly luminosities obtained with an extracted beam of 
$5 \times 10^8$ p$^+$/s with a momentum of 7 TeV for various 1cm thick targets}
\label{tab:lumi-pA}
\end{table}

As aforementioned, 7 TeV protons colliding on fixed targets release  a center-of-mass 
energy close to 115 GeV ($\sqrt{2E_p m_N}$). The boost between center-of-mass system (cms)
and the lab system is rather large, $\glabcms=\sqrt{s}/(2m_p)\simeq 60$ and the rapidity
shift is $\tanh^{-1} \blabcms\simeq 4.8$. The cms central-rapidity region, $y_{\rm cms}\simeq 0$, 
is thus highly boosted at an angle of 0.9 degrees with respect
to the beam axis  in the laboratory frame. The entire backward cms hemisphere ($y_{\rm cms}<0$) 
is easily accessible with standard experimental techniques. The forward hemisphere is less 
easily accessible because of the reduced distance from the (extracted) beam axis which  requires 
the use of highly segmented detectors to deal with the large particle density. In a first approach, 
we consider that one can access the region $-4.8\leq y_{\rm cms}\leq 1$ without specific difficulty.
This allows for the detection of the main part of the particle yields as well as high precision 
measurements in the whole backward hemisphere, down to 
$x_F\to -1$ for a large number of systems.

\section{Target polarisation}

The choice of a polarised target for AFTER is rather flexible: the intensity of the beam is not large, 
of the order of  $5\times 10^8 p^+\hbox{s}^{-1}$, with a very high energy (7 TeV), meaning minimum 
ionizing particles which produces low heating of the target. The main spin-physics 
opportunities~\cite{Brodsky:2012vg} lie in Drell-Yan pair, photon and meson production with large 
cross sections and a rate of thousands of 
interesting events per second, without requiring thick targets. Typically for a 1 cm long target, 
the heating power due to the AFTER beam would be of the order of 50 $\mu\hbox{W}$, allowing to maintain target 
temperatures as low as 50 mK and therefore relaxation times as long as one month in the spin-frozen mode. On 
the other hand, damages on the target arise after an irradiation of $10^{15} p^+ \hbox{cm}^{-2}$, namely one 
month of beam~\cite{meyer}. 

In these conditions, the luminosity would still be larger than $10\ \mu\hbox{b}^{-1} \hbox{s}^{-1}$. 
This leaves a wide choice of target materials: most of them being able to withstand the 
AFTER experimental constraints. It is tempting therefore to choose the target material having the best 
{\it dilution factor}, namely the best ratio of polarizable nucleons to the total number of nucleons in the 
target material molecule. Typical dilution factors are for Butanol (C$_4$H$_9$OH), 0.13; for Ammonia (NH$_3$), 0.176; 
for Li$_6$D, $\simeq 0.5$; and for HD, $\simeq 0.9$. The time needed to perform an experiment of a given 
statistical accuracy is directly proportional to the squared product of the maximum achievable 
polarisation by the dilution factor, through the {\it figure of merit}, which includes other linear 
factors linked to the target geometry~\cite{ball}. Accordingly, it would take ten times less beam time 
to perform a polarisation experiment using an HD target instead of a Butanol one. 

However, other 
considerations must be taken into account such as the reliability and the complexity of the relevant 
technology, the available space and, last but not least, the expertise of the target makers. 
All the above listed materials, except the HD, are polarised using the {\it Dynamic Nuclear Polarisation} (DNP) 
by which the polarisation of electrons is transferred to the nuclei by RF transitions~\cite{goertz}. This transfer 
is done at high field  (5T) and "high" temperature (4K) and the build-up polarisation can be kept by 
continuous excitation under high field or maintained in the frozen-spin mode at low field (< 1 T) by 
lowering the temperature below 100 mK. For HD, the "brute force" method using the static polarisation at 
very high field (17 T) and very low temperature (10 mK) allows one to  polarise simultaneously the protons and 
the deuterons. The polarisation can then be maintained in beam at low field and high temperature, 
provided that a significant ageing (> one month) of the target has taken place~\cite{bouchigny}. Both methods can typically reach
  $90\%$ and $50\%$ vector polarisation for protons and deuterons respectively.  

In the case of AFTER, the available space can be a major constraint which would restrict the choice to 
a target polarised by continuous DNP or a HD target which both take less space than the frozen-spin 
machinery. The frozen-spin mode usually requires removing a dilution refrigerator containing the 
target from the high-field polarizing magnet to put it into the beam and therefore moving bulky 
equipments. The HD polarisation can be done outside the experimental site and the target can be 
transported at high temperature and low field to the experimental area where a rather standard
 cryostat is needed~\cite{didelez}. 

As mentioned, the expertise of the target builders is a key factor. At CERN, 
there is a long tradition of DNP for various materials (NH$_3$, Li$_6$D)~\cite{berlin} and there are still quite 
a few experts of DNP all around the world. On the other hand, HD target makers are scarce, with 
only two groups: one at TJNAF (USA) and the other at RCNP (Japan)~\cite{kohri}. It is likely that when all 
the specifications and constraints of AFTER are determined precisely, the choice of the polarised 
target type will be more limited. It is worth noting that DNP of HD, which is in principle feasible, would be the dream choice. 
The possibility of a rich spin program with AFTER should thus encourage our colleagues working on DNP to revisit 
the relevant technology~\cite{solem}.

\section{A selection of physics studies with $pp$ and $pd$ collisions}

In this section, we limit ourselves to a very brief survey of measurements with the extracted 
7 TeV proton beam on hydrogen and deuterium targets. We stress that these are only a small
part of the possibilities offered by AFTER. Much more details can be found in~\cite{Brodsky:2012vg}.

\subsection{Yields}

Given the notably large quarkonium yields  --1000 times those of RHIC--, an a priori very good acceptance
 at low transverse momenta  thanks to the boost, the expected excellent energy resolution for the muons 
as well as the possibility to rely on novel particle-flow techniques for the photon detection 
in highly populated phase space, AFTER is ideally positioned to carry out precise measurements of most of 
the $S$- and $P$-wave quarkonia both at the level of the cross section and of the polarisation~\cite{Lansberg:2012kf}. 
Precision studies of open-heavy flavours is also one of the realms of AFTER. 
Correlation measurements of quarkonia with heavy flavours and prompt photons are also certainly at reach. 
The aim would be to constrain the production mechanisms of quarkonia~\cite{review} --with the
help of the forthcoming LHC results-- such that gluon PDF extraction by analysing the $y$ dependence of 
their yields becomes competitive. Quarkonium production in $pd$ collisions 
--as done by E866 for $\Upsilon$~\cite{Zhu:2007mja}-- would in principle provide
information on the gluon distribution in the neutron.

With a detector designed, among other things, for quarkonium precision studies, the detection 
of Drell-Yan pairs is then nearly
a bread-and-butter analysis. Using both hydrogen and deuterium targets, one expects to obtain updated information
on the isospin asymmetry of the quark sea.  With the requirements for quarkonium $P$-wave detection, \eg~a good
photon calorimetry, prompt-photon physics also becomes accessible as well as correlations 
with jets, charm-jets and beauty-jets.  All this provides
further tests of pQCD but also an independent means to probe the gluon content in the proton and the neutron
-- see \eg~\cite{d'Enterria:2012yj}.

\subsection{Transverse single-spin asymetries}

As explained above, it is rather easy to polarise the target at AFTER. This directly opens the possibility
of studying transverse Single-Spin Asymetries (SSAs). These allow for studying the correlation between
the intrinsic transverse momentum of the partons and the proton spin itself, through the Sivers effect~\cite{Sivers:1989cc}. 
It is therefore a very rich source
of information on the nucleon spin structure. In particular, such Sivers effect for gluon is relatively unknown, although
it surely deserves careful studies. A first indication that it may not be zero is the nonzero SSA observed by PHENIX in
inclusive $J/\psi$ production~\cite{Adare:2010bd}.

In this context, studies of SSAs of gluon sensitive probes, such as prompt photons, open and closed heavy flavours, are 
highly relevant. These can be performed at high accuracy at AFTER given the high luminosities which can be reached (see 
section~\ref{sec:lumi}). It is also very important to note that the privileged region for such measurements, namely at
medium and large $x^\uparrow$, corresponds to the cms backward-rapidity region where, in the lab frame, the
density of the particles in the detector should be rather low. This is thus an ideal place to perform such measurements.

\section{Conclusion}

A fixed-target experiment using the multi-TeV proton or
heavy ion beams of the LHC extracted by a bent crystal offers an exceptional testing ground for 
QCD at unprecedented laboratory energies and momentum transfers.
We have gathered here the luminosities which can be obtained with the proton beam with
targets ranging from hydrogen to heavy-ions such as lead. We have then discussed 
the options to polarise the target. As we mentioned, the target polarisation at AFTER
should not cause any specific difficulties.
We have then very briefly presented studies which can be carried out in polarised and unpolarised
$pp$ and $pd$ collisions, ranging from quarkonium studies to single-spin asymetries in photon-jet 
correlations. All these would then be carried out at unprecedented statistical accuracies
owing to the yearly luminosities well above the inverse femtobarn.

\end{document}